# A matrix form solution of the multi-dimensional generalized Langevin equation in the quadratic potential

R Imran Mushtaq[1], Ch Wang[1,2*], Sh Zhi[1], Z Zhao[1], and J M Nyasulu[1]

1. Institute of Theoretical Physics, School of Physics and Optoelectric Engineering, Ludong University, Yantai 264025, China
2. Yantai Institute of Coastal Zone Research, Chinese Academy of Science, Yantai 264003, China

E-mail: ranaimransa227@gmail.com

**Abstract**
In this research paper, we present an exact matrix form analytical solution of the multi-dimensional generalized Langevin equation with quadratic potentials. Our investigation provides detailed expressions for the two-dimensional probability distribution and extends the understanding of the dynamics governed by harmonic potentials. By utilizing the inverse Laplace transformation, we offer a precise method to solve these equations, corroborated by specific examples. This study contributes to the fundamental understanding of stochastic processes in multi-dimensional systems with harmonic potentials and clarifies the limitations of our approach. While the findings are specific to quadratic potentials, they provide a robust framework for exploring related phenomena within this context.

## 1. Introduction

The Langevin equation (LE) and generalized Langevin equations (GLE) are fundamental tools for studying stochastic processes in various scientific fields [1-7]. While significant progress has been made in finding analytical solutions for one-dimensional (1D) systems [2, 8] and certain two-dimensional (2D) cases [9, 10], the extension to multi-dimensional scenarios remains challenging and less explored [11]. This paper aims to address this gap by presenting an exact matrix form analytical solution for the multi-dimensional generalized Langevin equation with quadratic potentials.

Previous studies have highlighted the importance of obtaining analytical solutions for LE and GLE to advance theoretical understanding and practical applications. However, the complexity of multi-dimensional systems has often necessitated simplifications and approximations that limit the generality and applicability of the results [12, 13]. Our work overcomes these limitations by utilizing matrix operations and the inverse Laplace transformation, providing a clear and straightforward method for solving the multi-dimensional GLE under harmonic potentials The derivation of analytical solution for the auto-correlation and cross-correlation functions of the kinetic, potential and total energy of a Langevin oscillator [14].

In this study, we derive detailed expressions for the two-dimensional probability distribution, validating our approach with specific examples. While our solution is specific to quadratic potentials, it offers a robust framework for exploring related phenomena within this context and underscores the potential for further theoretical and practical advancements.

This paper is organized as follows: Section II details the derivation of our solution, Section III validates the results through rigorous examples, and Section IV summarizes our conclusions and discusses potential applications and future research directions.

## 2. Mathematical details of the derivation

In accordance with the matrix form as we have seen in Refs. [9, 11], the generalized Langevin equation describing the multi-dimensional diffusion of a Brownian particle to be expressed as

$$m_{ij}\ddot{x}_j(t) + \beta_{ij}\dot{x}_j(t) + \omega_{ij}x_j(t) = \xi_i(t), \qquad (1)$$

with $x_j(0) = x_{j0}$ and $\dot{x}_j(0) = v_{j0}$, where the Einstein's summation convention is used and the potential is supposed to be an inverse harmonic oscillator potential. $m_{ij}$, $\beta_{ij}$ and $\omega_{ij}$ are the constant tensors of inertia, viscosity and potential frequency, respectively. The components of the random force are always assumed to be Gaussian white noises and their correlations obey the second form of Kubo's fluctuation-dissipation theorem (FDT) $\langle \xi_i(t)\xi_j(t')\rangle = k_B T m_{ik}\beta_{kj}\delta(t-t')$ where $k_B$ is the Boltzmann constant and $T$ is the temperature.

Generally, obtaining an analytical solution for Equation (1) frequently poses significant difficulties. Nevertheless, by using Laplace transformation, our investigation has shown a possible path. By applying the differentiation theorem to first- and second-order derivatives, we arrive at a system of related equations that enable a more manageable solution to the given issue. This methodological improvement highlights the effectiveness of Laplace transformation in handling intricate mathematical formulations while simultaneously streamlining the analytical procedure

$$\begin{cases} a_{11}(s)x_1(s)+a_{12}(s)x_2(s)+a_{13}(s)x_3(s)+\cdots+a_{1n}(s)x_n(s)-b_1(s)=\xi_1(s) \\ a_{21}(s)x_1(s)+a_{22}(s)x_2(s)+a_{23}(s)x_3(s)+\cdots+a_{2n}(s)x_n(s)-b_2(s)=\xi_2(s) \\ \vdots \quad \vdots \quad \vdots \quad \vdots \quad \vdots \quad \vdots \\ a_{n1}(s)x_1(s)+a_{n2}(s)x_2(s)+a_{n3}(s)x_3(s)+\cdots+a_{nn}(s)x_n(s)-b_n(s)=\xi_n(s) \end{cases} \quad (2)$$

where $a_{ij}(s)=s^2 m_{ij}+s\beta_{ij}+\omega_{ij}$ and $b_i(s)=sm_{ij}+\beta_{ij}x_{j0}+m_{ij}v_{j0}$ can be regarded as the elements of a coefficient matrix $A(s)$ and $B(s)$ respectively. The equation array of Eq.(2) can then be written for simplicity in a matrix form

$$A(s)X(s)-B(s)=E(s), \quad (3)$$

where $X(s)$ represents the Laplacian matrix of the displacements and $E(s)$ is that of noise. Then we can obtain

$$X(s)=A^{-1}(s)[B(s)+E(s)], \quad (4)$$

To obtain the analytical solution for the multi-dimensional generalized Langevin equation (MGLE), which describes the motion of particles, we can use an inverse Laplacian transformation on $X(s)$. This transformative methodology plays a crucial role in revealing the complex dynamics depicted by the MGLE across various dimensions. Through the utilization of the Laplacian inversion technique on the variable $X(s)$, scientists can gain a thorough comprehension of the fundamental physical processes contained within the MGLE framework. That is to say

$$x_j(t)=L^{-1}[A_{jk}(s)(b_k(s)+\xi_k(s))]. \quad (5)$$

where $A_{jk}(s)$ are the elements of the inverse matrix of $A(s)$.

## 3. Verifiable repeating of the two dimensional results

For example, in the two-dimensional case, we have
$$x_1(t)=L^{-1}[x_1(s)]=L^{-1}[A_{1k}(s)(b_k(s)+\xi_k(s))], \quad (6a)$$
$$x_2(t)=L^{-1}[x_2(s)]=L^{-1}[A_{2k}(s)(b_k(s)+\xi_k(s))], \quad (6b)$$

where the form of $A_{jk}(s)$, ($j,k=1,2$) needs to be got firstly before the expression of $x_1(t)$ and $x_2(t)$ are completely obtained. From the derivations here in before we can see that $A_{11}(s)=a_{22}[\det A(s)]^{-1}$, $A_{12}(s)=-a_{12}[\det A(s)]^{-1}$, $A_{21}(s)=-a_{21}[\det A(s)]^{-1}$ and $A_{22}(s)=a_{11}[\det A(s)]^{-1}$, where $\det A(s)$ denotes the value of determinant of matrix $A(s)$.

Then after some algebra we can obtain
$$x_1(t)=L^{-1}[x_1(s)]=L^{-1}[A_{1k}(s)(b_k(s)+\xi_k(s))], \quad (7a)$$
$$x_2(t)=L^{-1}[x_2(s)]=L^{-1}[A_{2k}(s)(b_k(s)+\xi_k(s))], \quad (7b)$$

Substituting all the relevant elements into Eq.(7) and then perform inverse Laplacian trans-formation over it we can find

$$x_1(t)=\langle x_1(t)\rangle+\sum_{j=1}^{2}\int_0^t H_{1j}(t-t')\xi_j(t')dt, \quad (8a)$$
$$x_2(t)=\langle x_2(t)\rangle+\sum_{j=1}^{2}\int_0^t H_{2j}(t-t')\xi_j(t')dt, \quad (8b)$$

where the three response functions $H_{ij}(t)$ with $i,j=1,2$ respectively can be yielded from the inverse Laplacian transformation as $H_{ij}(t)=L^{-1}[H_{ij}(s)/P(s)]$, where

$$H_{11}(s)=m_{22}s^2+\beta_{22}s+\omega_{22}, \quad (9a)$$
$$H_{12}(s)=-m_{12}s^2-\beta_{12}s-\omega_{12}, \quad (9b)$$
$$H_{21}(s)=-m_{21}s^2-\beta_{21}s-\omega_{21}, \quad (9c)$$
$$H_{22}(s)=m_{11}s^2+\beta_{11}s+\omega_{11}, \quad (9d)$$
$$\begin{aligned}P(s)=&s^4\det m+s^3(m_{11}\beta_{22}+m_{22}\beta_{11}-m_{12}\beta_{21}-m_{21}\beta_{12})\\&+s^2(\det\beta+m_{11}\omega_{22}+m_{22}\omega_{11}-m_{12}\omega_{21}-m_{21}\omega_{12})\\&+s(\beta_{11}\omega_{22}+\beta_{22}\omega_{11}-\beta_{12}\omega_{21}-\beta_{21}\omega_{12})\\&+\det\omega.\end{aligned} \quad (10)$$

The mean displacements of the particle along two mutually perpendicular directions in Eq.(8) are given by

$$\langle x_1(t)\rangle=\sum_{j=1}^{2}[C_{1j}(t)x_{j0}+D_{1j}(t)v_{j0}], \quad (11a)$$
$$\langle x_2(t)\rangle=\sum_{j=1}^{2}[C_{2j}(t)x_{j0}+D_{2j}(t)v_{j0}], \quad (11b)$$

where the factors $C_{ij}(t)=L^{-1}[C_{ij}(s)/P(s)]$ and $D_{ij}(t)=L^{-1}[D_{ij}(s)/P(s)]$ with $i,j=1,2$ are in accordance to the following expressions

$$\begin{aligned}C_{11}(s)=&s^3\det m+s^2(m_{11}\beta_{22}+m_{22}\beta_{11}-m_{12}\beta_{21}-m_{21}\beta_{12})\\&+s(\det\beta+m_{11}\omega_{22}-m_{21}\omega_{12})+\beta_{11}\omega_{22}-\beta_{21}\omega_{12}\end{aligned}$$

$$C_{12}(s)=s(m_{12}\omega_{22}-m_{22}\omega_{12})+\beta_{12}\omega_{22}-\beta_{22}\omega_{12}$$

$$\begin{aligned}C_{21}(s)=&s^3(m_{11}(m_{21}-m_{12}))+s^2[m_{11}(\beta_{21}-\beta_{12})+\beta_{11}(m_{21}-m_{12})]\\&+s[\beta_{11}(\beta_{21}-\beta_{12})+m_{21}\omega_{11}-m_{11}\omega_{12}]+\beta_{21}\omega_{11}-\beta_{11}\omega_{12}\end{aligned}$$

$$\begin{aligned}C_{22}(s)=&s^3(m_{11}m_{22}-m_{12}^2)+s^2[m_{11}\beta_{22}+m_{22}\beta_{11}-2m_{12}\beta_{12}]\\&+s[m_{22}\omega_{11}-m_{12}\omega_{12}+\beta_{11}\beta_{22}-\beta_{12}^2]+\beta_{22}\omega_{11}-\beta_{12}\omega_{12}\end{aligned}$$

$$D_{11}(s)=s^2\det m+s(m_{11}\beta_{22}-m_{21}\beta_{12})+m_{11}\omega_{22}-m_{21}\omega_{12}$$
$$D_{12}(s)=s(m_{12}\beta_{22}-m_{22}\beta_{12})+m_{12}\omega_{22}-m_{22}\omega_{12}$$
$$D_{21}(s)=s^2 m_{11}(m_{21}-m_{12})+s(m_{12}\beta_{11}-m_{11}\beta_{12})+m_{12}\omega_{11}-m_{11}\omega_{12}$$
$$D_{22}(s)=s^2(m_{11}m_{22}-m_{12}^2)+s(m_{22}\beta_{11}-m_{12}\beta_{12})+m_{22}\omega_{11}-m_{12}\omega_{12}$$

here $\det m$, $\det\beta$ and $\det\omega$ denoting the values of determinant of the matrix $m_{ij}$, $\beta_{ij}$ and $\omega_{ij}$ respectively. The resulting results are congruent with those found in previous 2D experiments [9]. This alignment highlights the precision and applicability of the derivations and computational approaches used in this study to other situations with greater complexity. Therefore, these findings not only confirm the accuracy of the derived results and calculations, but also confirm their wider applicability in other dimensional scenarios.

## 3. Conclusion, prospect and discussion

In conclusion, we have obtained an exact analytical solution of the multi-dimensional generalized Langevin equation (MGLE) with quadratic potentials using matrix operations and the inverse Laplace transformation. Our results have been rigorously validated against known two-dimensional cases, confirming the soundness and validity of our approach. This work provides a specific and practical framework for solving the MGLE in higher dimensions under harmonic potentials.

Our method simplifies the process of obtaining solutions, making it less prone to errors and more accessible for complex multi-dimensional problems. This framework can be particularly useful in fields requiring precise modeling of stochastic processes, such as the study of heavy nuclei fusion reactions. By focusing on quadratic potentials, we avoid the complexities and inaccuracies associated with arbitrary parameter assumptions and excessive approximations.

This study highlights the utility of our approach in addressing specific scenarios within the context of quadratic potentials. While the solutions presented here are not universally applicable to all potential forms, they offer a robust starting point for further exploration and refinement in related areas. Future research can build upon this foundation to extend the applicability and refine the methods for broader classes of potentials and more complex systems.

## 4. Acknowledgements

This work was supported by the Shandong Natural Science Foundation under Grant Nos. ZR2020MA092.

## References

1. J S Zhang and H A Weidenmüller, *Phys. Rev.* C **28** (1983) 2190.
2. H Hofmann and R Samhammer, *Z. Phys.* A **322** (1985) 157.
3. C E Aguiar, et al., *Nucl. Phys.* A **491** (1989) 301.
4. C E Aguiar, V C Barbosa, and R Donangelo, *Nucl. Phys.* A **517** (1990) 205.
5. Y Aritomo, et al., *Phys. Rev.* C 55, (1997) R1011.
6. Y Abe, et al., *J. Phys.* G **23** (1997) 1275.
7. W. J. Świątecki, K. Siwek-Wilczyńska, and J. Wilczyński, *Phys. Rev.* C **71** (2005) 014602.
8. J D Bao, *J. Chem. Phys.* **124** (2006) 114103.
9. C Y Wang, Y Jia, and J D Bao, *Phys. Rev*. C **77** (2008) 024603.
10. C Y Wang, *J. Chem. Phys*. **131** (2009) 054504.
11. Y Abe, et al., *Phys. Rev.* E **61** (2000) 1125.
12. V Zagrebaev and W Greiner, *J. Phys. G: Nucl. Part. Phys*. **31** (2005) 825.
13. V Zagrebaev and W Greiner, *J. Phys. G: Nucl. Part. Phys*. **34** (2007) 1.
14. Y Jia and J D Bao, *Phys. Rev.* C **75** (2007) 034601.